\documentclass{article}

\usepackage{arxiv}

\usepackage[utf8]{inputenc} % allow utf-8 input
\usepackage[T1]{fontenc}    % use 8-bit T1 fonts
\usepackage{hyperref}       % hyperlinks
\usepackage{url}            % simple URL typesetting
\usepackage{booktabs}       % professional-quality tables
\usepackage{amsfonts}       % blackboard math symbols
\usepackage{nicefrac}       % compact symbols for 1/2, etc.
\usepackage{microtype}      % microtypography
\usepackage{lipsum}		% Can be removed after putting your text content
\usepackage{graphicx}
\usepackage{cite}
\usepackage{doi}
\usepackage{amsmath}
\usepackage{multirow}

\title{Novel-view X-ray Projection Synthesis through Geometry-Integrated Deep Learning}

\author{
\href{https://orcid.org/0009-0001-0905-3833}{\includegraphics[scale=0.06]{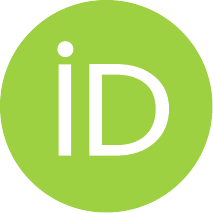}\hspace{1mm}Daiqi Liu}\thanks{Corresponding author: daiqi.deutschfau.liu@fau.de} \quad
\href{https://orcid.org/0000-0001-9602-6237}{\includegraphics[scale=0.06]{orcid.pdf}\hspace{1mm}Fuxin Fan} \quad
\href{https://orcid.org/0000-0000-0000-0000}{\includegraphics[scale=0.06]{orcid.pdf}\hspace{1mm}Andreas Maier} \\
Pattern Recognition Lab, Friedrich-Alexander-Universität Erlangen-Nürnberg \\
\texttt{\ daiqi.deutschfau.liu@fau.de, fuxin.fan@fau.de, andreas.maier@fau.de}
}
\date{}

\renewcommand{\headeright}{Liu, Fan \& Maier}
\renewcommand{\undertitle}{}
\renewcommand{\shorttitle}{DL-GIPS Framework }

\hypersetup{
pdftitle={Novel-view X-ray Projection Synthesis through Geometry-Integrated Deep Learning},
pdfauthor={Daiqi Liu, Fuxin Fan, Andreas Maier},
}

\begin{document}
\maketitle

\begin{abstract}
X-ray imaging plays a crucial role in the medical field, providing essential insights into the internal anatomy of patients for diagnostics, image-guided procedures, and clinical decision-making. Traditional techniques often require multiple X-ray projections from various angles to obtain a comprehensive view, leading to increased radiation exposure and more complex clinical processes. This paper explores an innovative approach using the DL-GIPS model, which synthesizes X-ray projections from new viewpoints by leveraging a single existing projection. The model strategically manipulates geometry and texture features extracted from an initial projection to match new viewing angles. It then synthesizes the final projection by merging these modified geometry features with consistent texture information through an advanced image generation process. We demonstrate the effectiveness and broad applicability of the DL-GIPS framework through lung imaging examples, highlighting its potential to revolutionize stereoscopic and volumetric imaging by minimizing the need for extensive data acquisition. The source code are publicly available at \url{https://github.com/DaE-plz/Re-GIPS} for further research and development.
    
\end{abstract}

\section{Introduction}
X-ray imaging plays a vital role in visualizing internal patient structures for diagnostics, interventions, and clinical decision-making. Traditional X-ray imaging often requires multiple projections from various angles to comprehensively capture anatomical details. This approach increases radiation exposure and complicates clinical workflows. Computational image synthesis has emerged as a promising alternative to reduce the cost and complexity of obtaining multi-view projections while minimizing radiation \cite{3639-01}. Despite its potential, this area remains underexplored.

Recent advances in deep learning have driven progress in fields such as image reconstruction and recognition \cite{3639-02, 3639-03, 3639-07}. In particular, computer vision has leveraged deep learning to achieve novel view synthesis \cite{3639-04, 3639-05, 3639-06}. However, X-ray imaging presents unique challenges due to its reliance on penetrating rays rather than reflected light, necessitating specialized solutions.

This paper introduces the DL-GIPS framework, which combines geometric transformations with deep learning to synthesize X-ray projections from new viewpoints. DL-GIPS extracts geometry and texture features from initial projections, aligns them with target angles through geometric transformations, and synthesizes high-quality projections that respect the physical properties of X-ray imaging while preserving textural continuity.

The contributions of this work are as follows: (1) reproducing and optimizing the DL-GIPS network, (2) proving the model's robustness, and (3) validating the framework through extensive experiments on lung imaging datasets, demonstrating its effectiveness in both one-to-one and multi-to-multi projection synthesis scenarios.

\section{Materials and methods}
\subsection{DL-GIPS framework}

\begin{figure}[h]
  \centering
  \includegraphics[width=0.8\linewidth]{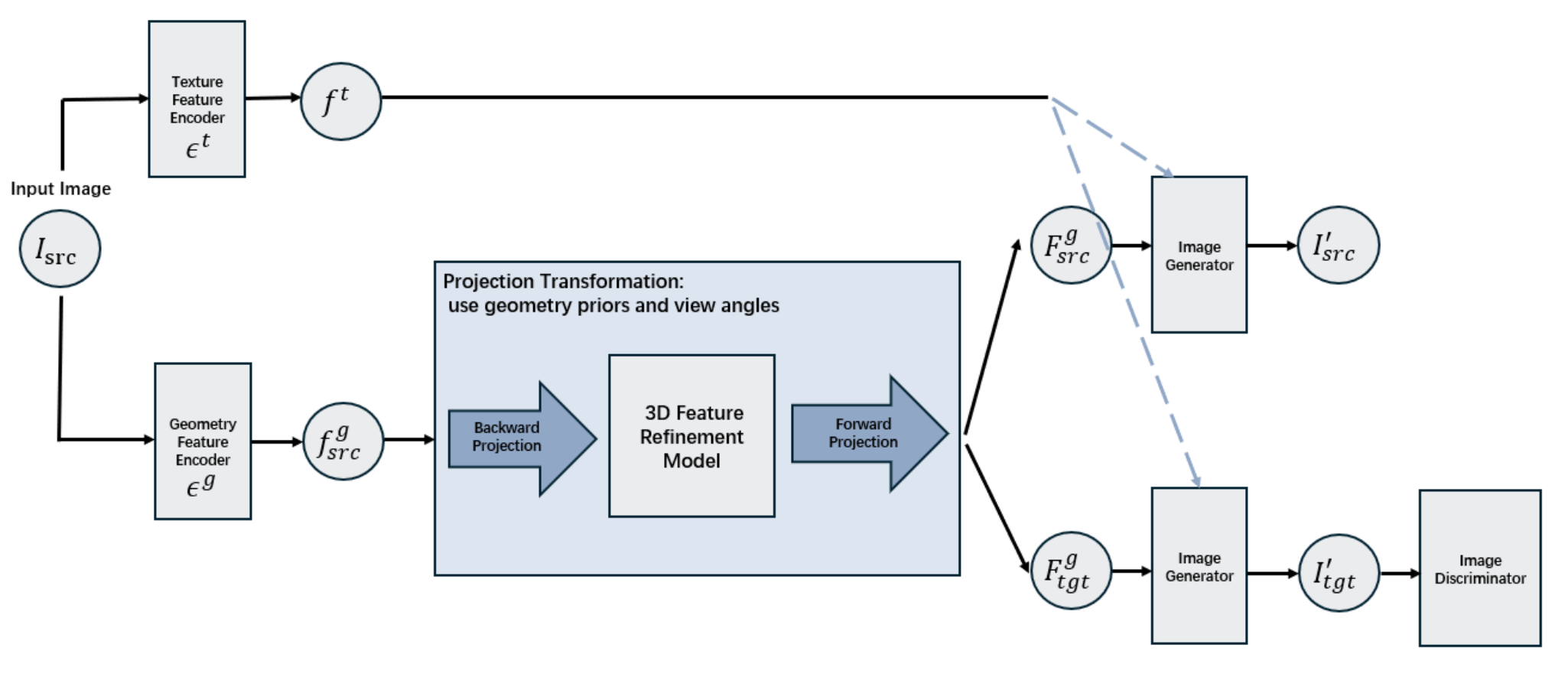}
  \caption{Illustration of the DL-GIPS. The pipeline contains texture and geometry feature encoders, projection transformation and image generator.}
  \label{dlgipsframework}
\end{figure}

As illustrated in Figure.~\ref{dlgipsframework}, The DL-GIPS framework is designed to synthesize X-ray projections from novel viewpoints using a single or multiple existing projections. The framework consists of four primary components: feature extraction, geometric transformation, image generation, and image discriminator.

\subsubsection{Feature Extraction}
Feature extraction is the first step in the DL-GIPS framework, where geometry and texture features are disentangled from the input X-ray projection(s). Two separate encoders are constructed to learn the geometry and texture features from the input projection respectively, denoted as geometry feature encoder $\varepsilon^g$, and texture feature encoder $\varepsilon^t$. Given the source-view image as $I_{\text{src}}$, the feature encoding process is denoted as follows:
\begin{equation}
f_{\text{src}}^g = \varepsilon^g(I_{\text{src}}), f_{\text{src}}^t = \varepsilon^t(I_{\text{src}})
\end{equation} 

\subsubsection{Geometric Transformation}

The geometry features extracted from the source-view projection are first backward-projected onto a 3D feature volume, which represents the semantic features of the underlying subject in 3D space. Next, the 3D feature volume is forward-projected onto the target image plane according to the cone-beam geometry and target view angle. In this way, the geometry features extracted from the source-view projection are transformed into the corresponding features for the target view.

When the input projection(s) is sparse (e.g., a single source view), we introduce a 3D feature refinement model, to inpaint the 3D feature volume after the back-projection operation. The entire projection transformation process can be expressed as:
\begin{equation}
F_{\text{src}}^g, F_{\text{tgt}}^g = P^f M \circ P^b(f_{\text{src}}^g)
\end{equation}

where we denote the transformed features after forward and back- ward projection as $F_{\text{src}}^g$ and $F_{\text{tgt}}^g$. $P^b$ and $P^f$ represent the forward pro- jection and backward projection operators. $M$ represents the 3D feature refinement network.

\subsubsection{Image Generation}
The image generation module combines the transformed geometry features and the texture features to synthesize the target-view projection. The generator, denoted as $g$, outputs both the target-view and source-view projections to ensure consistency:
\begin{equation}
I'_{\text{src}} = g(F^g_{\text{src}}, f^t), I'_{\text{tgt}} = G(F^g_{\text{tgt}}, f^t)
\end{equation}

\subsubsection{Image Discriminator}
To enhance the realism of the synthesized projections, an adversarial training approach is employed with a multi-scale image discriminator $D$. The discriminator distinguishes real projections from synthesized ones across multiple scales, enforcing alignment with the distribution of real data.

\subsection{Dataset}
The experiments utilized the Lung Image Database Consortium and Image Database Resource Initiative (LIDC-IDRI) dataset, comprising 1018 thoracic 3D CT images from various patients \cite{3639-08}. After filtering out irrelevant information, the dataset was reduced to 819 patient data samples. All CT images were resampled to a uniform resolution of 1 mm in the z-axis and resized to a consistent image size of $128 \times 128$ in the xy-plane. Additionally, after volume padding, the 3D volume dimensions were adjusted to $(500, 128, 128)$.

\begin{figure}
    \centering
    \includegraphics[width=0.8\textwidth]{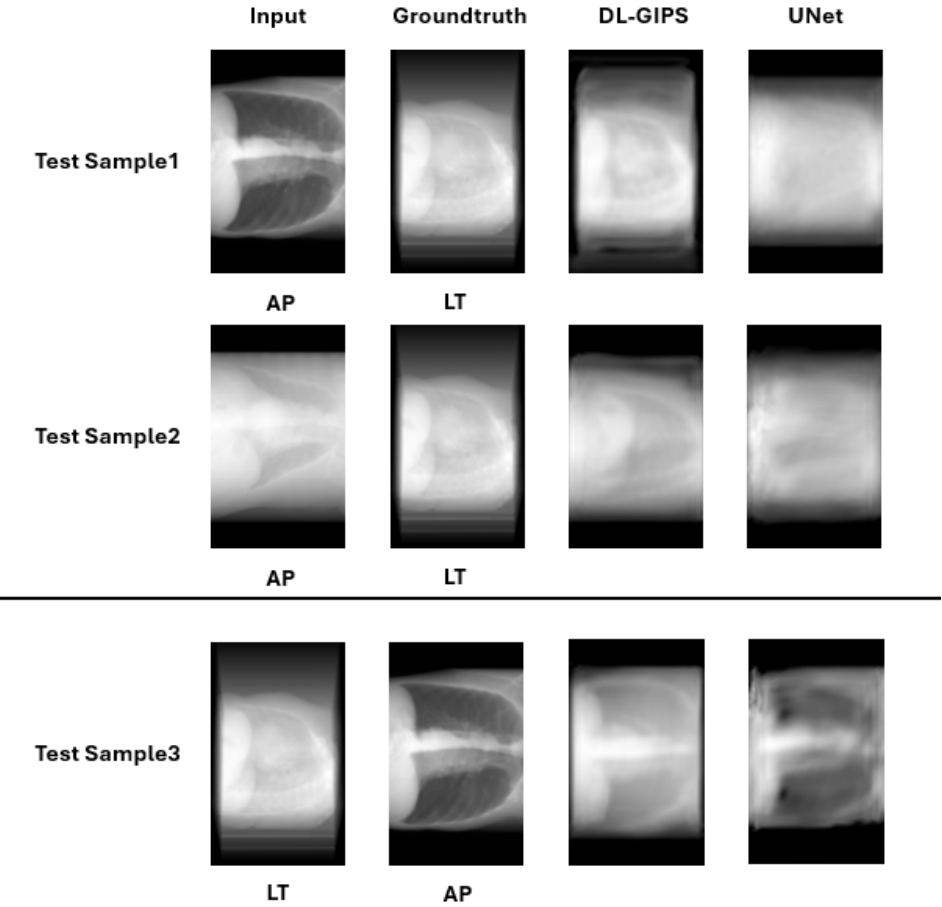} 
    \caption{Illustration of one-to-one view synthesis experiments.}
    \label{fig:one_to_one}
\end{figure}

To generate X-ray projections at different angles, digital reconstruction radiographs (DRRs) were derived from the 3D CT images. The cone-beam geometry of the projection operation was defined based on a clinical on-board cone-beam CT system for radiation therapy. Each 2D X-ray projection had dimensions of $180 \times 300$. The intensity values of all 2D X-ray projection images were normalized to the range $[0, 1]$. For experimentation, 80\% of the dataset (656 samples) was randomly selected for training and validation, while the remaining 20\% (163 samples) were reserved for testing.

\subsection{Training Loss}

Consistency Loss ensures that the extracted features from the synthesized image can reconstruct the input image while preserving the geometric features of the synthesized projection, ensuring they remain consistent with the previously transformed geometry features. Reconstruction loss measures the difference between the synthesized projection and the ground truth projection, ensuring that the transformed geometry features can accurately generate the novel view while preserving consistency in the source view. Adversarial loss ensures that the synthesized projections resemble real projections by training a discriminator to distinguish between real and generated images, encouraging the generator to produce more realistic results. 

The feature encoders, image generators, and image discriminators are jointly trained to optimize the total loss as follows:

\begin{equation}
L= \lambda_{cyc} L_{cyc} + \lambda_{rec} L_{rec} + \lambda_{adv} L_{adv}
\end{equation}

where $\lambda_{cyc}$, $\lambda_{rec}$, $\lambda_{adv}$ are the hyper-parameters to balance the different parts of the total loss.

\section{Results}

Two experimental settings were designed to validate the proposed model under different view synthesis scenarios: one-to-one view synthesis and multi-to-multi view synthesis. 

As shown in Figure~\ref{fig:one_to_one}, in one-to-one view synthesis experiments, we conducted experiments to generate either the anteroposterior (AP) projection from the lateral (LT) projection or the LT projection from the AP projection. 

\begin{figure}[hb]
    \centering
    \includegraphics[width=0.8\textwidth]{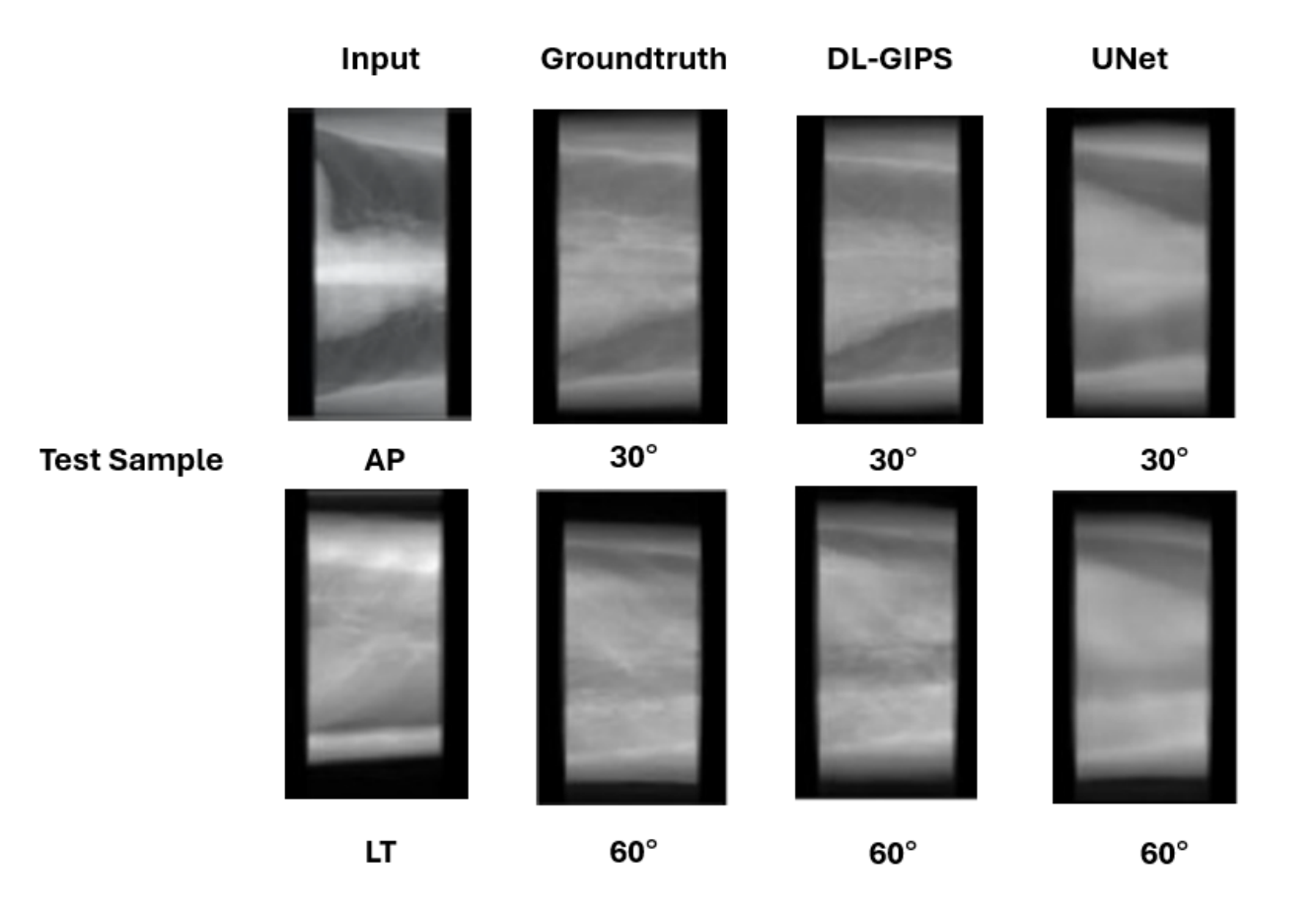} 
    \caption{Illustration of multi-to-multi view synthesis experiments.}
    \label{fig:two_to_two}
\end{figure}

As shown in Figure~\ref{fig:two_to_two}, In multi-to-multi view synthesis experiments, the model was trained to generate projections at 30° and 60° from both the AP (0°) and LT (90°) projections. 

The corresponding quantitative results averaged across all the testing data are reported in Table~\ref{tab:view_synthesis_results} for both "AP $\to$ LT", "LT $\to$ AP", and "Multi $\to$ Multi". The evaluation metrics include mean absolute error (MAE), normalized root mean squared error (RMSE), structural similarity (SSIM), and peak signal-to-noise ratio (PSNR). DL-GIPS achieves MAE values of 0.052 for "AP $\to$ LT" and 0.051 for "LT $\to$ AP", compared to UNet's 0.078 and 0.073, respectively \cite{3639-09}. The PSNR is also improved, with DL-GIPS reaching 19.46 for "AP $\to$ LT" and 20.17 for "LT $\to$ AP", surpassing UNet's 16.43 and 18.58. In the multi-to-multi view synthesis scenario, DL-GIPS again outperforms UNet. DL-GIPS achieves a lower MAE of 0.017, RMSE of 0.12, and higher SSIM (0.81) and PSNR (23.53) compared to UNet’s 0.021, 0.13, 0.79, and 21.46, respectively.

\begin{table}[ht]
    \centering
    \caption{Results of novel-view projection synthesis.}
    \label{tab:view_synthesis_results}
    \renewcommand{\arraystretch}{1.2}
    \setlength{\tabcolsep}{8pt} 
    \begin{tabular}{l|cccc|cccc}
        \hline
        \multirow{2}{*}{Methods} & \multicolumn{4}{c|}{UNet} & \multicolumn{4}{c}{DL-GIPS} \\
        \cline{2-9}
        & MAE & RMSE & SSIM & PSNR & MAE & RMSE & SSIM & PSNR \\
        \hline
        AP $\rightarrow$ LT & 0.078 & 0.362 & 0.851 & 16.431 & 0.052 & 0.272 & 0.862 & 19.458 \\
        LT $\rightarrow$ AP & 0.073 & 0.341 & 0.871 & 18.583 & 0.051 & 0.256 & 0.893 & 20.167 \\
        Multi$\rightarrow$Multi & 0.031 & 0.132 & 0.793 & 21.461 & 0.034 & 0.116 & 0.814 & 23.525 \\
        \hline
    \end{tabular}
\end{table}

\section{Discussion}

In this work, we reproduced the DL-GIPS network for novel-view synthesis of X-ray projections, which successfully generates X-ray projections at target-view angles given source-view projections. The synthesized projections can be applied in various clinical scenarios, these applications could reduce radiation exposure and improve clinical workflows, especially for pediatric or pregnant patients.

The DL-GIPS model integrates geometry priors from the X-ray imaging system to relate source and target view angles. However, the current method's computational efficiency remains suboptimal due to the geometry transformation, with inference times for the DL-GIPS model higher than traditional deep learning models like UNet. In the experiment setting of one-to-one projec- tion synthesis, the inference time of one data sample for differ- ent methods are around: 0.04 s, 0.56 s for UNet and DL-GIPS models respectively.

In the future, we can explore the use of real clinical datasets for more accurate validation, or alternatively, utilize Deep DRR to simulate high-quality X-ray projections with more realistic features. This will help enhance the generalization of our model to real-world applications.

\section*{Acknowledgements}
The authors gratefully acknowledge the scientific support and HPC resources provided by the Erlangen National High Performance Computing Center (NHR@FAU) of the Friedrich-Alexander-Universität Erlangen-Nürnberg (FAU). The hardware is funded by the German Research Foundation (DFG).

\bibliographystyle{unsrt}

 %%% Uncomment this line and comment out the ``thebibliography'' section below to use the external .bib file (using bibtex) .


\begin{thebibliography}{1}

\bibitem{3639-01}
Wei Zhao, Liyue Shen, Md~Tauhidul Islam, Wenjian Qin, Zhicheng Zhang, Xiaokun Liang, Gaolong Zhang, Shouping Xu, and Xiaomeng Li.
\newblock Artificial intelligence in image-guided radiotherapy: a review of treatment target localization.
\newblock {\em Quantitative imaging in medicine and surgery}, 11(12):4881, 2021.

\bibitem{3639-02}
Liyue Shen, Wei Zhao, Dante Capaldi, John Pauly, and Lei Xing.
\newblock A geometry-informed deep learning framework for ultra-sparse 3d tomographic image reconstruction.
\newblock {\em Computers in Biology and Medicine}, 148:105710, 2022.

\bibitem{3639-03}
Kaiming He, Xiangyu Zhang, Shaoqing Ren, and Jian Sun.
\newblock Deep residual learning for image recognition.
\newblock In {\em Proceedings of the IEEE conference on computer vision and pattern recognition}, pages 770--778, 2016.

\bibitem{3639-07}
Alex Krizhevsky, Ilya Sutskever, and Geoffrey~E Hinton.
\newblock Imagenet classification with deep convolutional neural networks.
\newblock {\em Advances in neural information processing systems}, 25, 2012.

\bibitem{3639-04}
SM~Ali Eslami, Danilo Jimenez~Rezende, Frederic Besse, Fabio Viola, Ari~S Morcos, Marta Garnelo, Avraham Ruderman, Andrei~A Rusu, Ivo Danihelka, Karol Gregor, et~al.
\newblock Neural scene representation and rendering.
\newblock {\em Science}, 360(6394):1204--1210, 2018.

\bibitem{3639-05}
Vincent Sitzmann, Michael Zollh{\"o}fer, and Gordon Wetzstein.
\newblock Scene representation networks: Continuous 3d-structure-aware neural scene representations.
\newblock {\em Advances in neural information processing systems}, 32, 2019.

\bibitem{3639-06}
Ben Mildenhall, Pratul~P Srinivasan, Matthew Tancik, Jonathan~T Barron, Ravi Ramamoorthi, and Ren Ng.
\newblock Nerf: Representing scenes as neural radiance fields for view synthesis.
\newblock {\em Communications of the ACM}, 65(1):99--106, 2021.

\bibitem{3639-08}
G~Samuel.
\newblock The lung image database consortium (lidc) and image database resource initiative (idri): A completed reference database of lung nodules on ct scans.
\newblock {\em Medical physics}, 38:2, 2011.

\bibitem{3639-09}
Olaf Ronneberger, Philipp Fischer, and Thomas Brox.
\newblock U-net: Convolutional networks for biomedical image segmentation.
\newblock In {\em Medical image computing and computer-assisted intervention--MICCAI 2015: 18th international conference, Munich, Germany, October 5-9, 2015, proceedings, part III 18}, pages 234--241. Springer, 2015.

\end{thebibliography}




% \end{thebibliography}

\end{document}